\documentclass[]{spie}  %>>> use for US letter paper
 \pdfoutput=1
\usepackage{amsmath,amsfonts,amssymb}
\usepackage{graphicx}
\usepackage[colorlinks=true, allcolors=blue]{hyperref}
\usepackage[]{xfrac}

\title{Analysis techniques for complex-field radiation pattern measurements}

\author[a]{Kristina K. Davis}
\author[b]{Willem Jellema}
\author[c]{Stephen J. C. Yates}
\author[d]{Christopher E. Groppi}
\author[e]{Jochem J. A. Baselmans}
\author[b,c]{Andrey M. Baryshev}
\affil[a]{Department of Physics, University of California, Santa Barbara, CA, USA}
\affil[b]{Kapteyn Astronomical Institute, University of Groningen, 9700 AV Groningen, NL;}
\affil[c]{SRON Netherlands Institute for Space Research, Landleven 12, 9747 AD Groningen, NL}
\affil[d]{Arizona State University, 781 E. Terrace Drive, Tempe, AZ, USA}
\affil[e]{SRON Netherlands Institute for Space Research, Sorbonnelaan 2, 3584 CA Utrecht, NL}

\authorinfo{Further author information: (Send correspondence to K.K.D) \\K.K.D. E-mail: kdavis32@asu.edu\\  S.J.C.Y. E-mail: S.Yates@sron.nl}

\begin{document} 
\maketitle

\begin{abstract}
Complex field measurements are increasingly becoming the standard for state-of-the-art astronomical instrumentation. Complex field measurements have been used to characterize a suite of ground, airborne, and space-based heterodyne receiver missions \cite{Tong1994-NFbeams2D-submm,Tong2003-NFbeams1THz,Carter:ISSTT02,Jellema-PhD2015,Naruse2009-ALMAb8NFbeams,ALMAband9}, and a description of how to acquire coherent field maps for direct detector arrays was demonstrated in Davis et. al. 2017 \cite{Davis2017-vector-beam}. This technique has the ability to determine both amplitude and phase radiation patterns from individual pixels on an array for direct comparison to optical simulations. Phase information helps to better characterize the optical performance of the array (as compared to total power radiation patterns) by constraining the fit in an additional plane \cite{Jellema-PhD2015}. This is a powerful technique to diagnose optical alignment errors through the optical system, as a complex field scan in an arbitrary plane can be propagated either forwards or backwards through optical elements to arbitrary planes along the principal axis. Complex radiation patterns have the advantage that the effects of optical standing waves and alignment errors between the scan system and the instrument can be corrected and removed during post processing. 

Here we discuss the mathematical framework used in an analysis pipeline developed to process complex field radiation pattern measurements. This routine determines and compensates misalignments of the instrument and scanning system. We begin with an overview of Gaussian beam formalism and how it relates to complex field pattern measurements. Next we discuss a scan strategy using an offset in $z$ along the optical axis that allows first-order optical standing waves between the scanned source and optical system to be removed in post-processing. Also discussed is a method by which the co- and cross-polarization fields can be extracted individually for each pixel by rotating the two orthogonal measurement planes until the signal is the co-polarization map is maximized (and the signal in the cross-polarization field is minimized).  We detail a minimization function that can fit measurement data to an arbitrary beam shape model. We conclude by discussing the angular plane wave spectral (APWS) method for beam propagation, including the near-field to far-field transformation. 
\end{abstract}

\section{INTRODUCTION}
\label{sec:intro}  % \label{} allows reference to this section

Astronomers design instruments to meet specific sensitivity, field of view, and resolution requirements suitable to their scientific purpose. Once assembled, the instrument must be tested  to determine how well it matches the designed performance. There are many figures of  merit to characterize the optical performance of an instrument, but here we focus on measuring the instrument's radiation (beam) pattern as a way to check system alignment and optical efficiency. The optical properties of the individual components must be calibrated to model the behavior of the complete assembly, and careful alignment during system integration is necessary to ensure maximum sensitivity and resolution. 

The main figures of merit a beam pattern can measure are:
\begin{enumerate}
\item A radiation pattern measurement  is analogous to measuring a beam's ‘shape’ and 'size'. The beam 'shape' or spatial pattern is used to calibrate the absolute brightness of a point source depending on its position within the beam, and the beam size determines the resolution of the telescope. Irregularities in the shape of the beam can be used to diagnose misalignments between elements of the optical system or manufacturing errors of the receiver components.
\item The beam pattern is a way to measure the degree of coupling to a point source, which is a measure of the sensitivity of the instrument. The efficiency of the system is referred to as the 'throughput' of the system, measuring what fraction of photons that hit the primary mirror make it to the detector. Maximum sensitivity is achieved when there are no optical losses through the instrument. Beam patterns on the ground or in the lab can diagnose focusing errors, reflections, or misalignments that introduce losses into the system.   
\item A beam pattern measurement can also determine the pointing direction of the optical elements, referred to as the boresight angle of the beam. The boresight angle of the detector is measured relative to the optical axis of the telescope's primary mirror, most often measured using spherical coordinates with an origin at the focal plane. Typically, instruments are designed to have a boresight angle of zero, though this may not be the case for array instruments, off-axis optical configurations, or special circumstances where a beam does not fill the primary mirror. It is important to ensure proper alignment to ensure the telescope is aimed at the target. 
\item Lastly, for array instruments, the beam pattern measurement is used to determine the imaging properties of the array. The beam pattern measurements can look at beam shape, sensitivity, and pointing direction as a function of pixel position. One important factor is how well the beams overlap as they appear on the sky, also referred to as the filling factor of the image. This determines whether a single pointing will fully capture and astronomical image or if the instrument will have to shift or dither to fully sample an image.  
\end{enumerate}

A radiation pattern measurement can be performed in a laboratory setting or using on-sky calibration targets during astronomical observations. For astronomical cameras, the radiation pattern is usually measured with the camera in the receiving mode, though for special circumstances it may be more convenient to measure parts of the optical system in a transmitting mode. Both the thermal and optical environment around the camera can affect the measured radiation pattern, so for accurate results most measurements are made in a carefully controlled environment. For example, in the THz frequency regime, absorbing materials are not widely available, so scattering materials are usually coated on surfaces surrounding the optical system. Scattering dilutes the incoming radiation and reduces the fraction of light that is eventually scattered back into the optical path.

To take a beam pattern measurement, either a source or the receiver element is mounted on a mechanical scanning system, and the instrument response is recorded as a function of the scanner position. Most often, an emitting source probe is mounted on a planar scanning system and scans are made in a Cartesian coordinate grid, though for some applications cylindrical or spherical scans are used.  Nyquist sampling theorem \cite{LANDAU1967-Nyquist-vs-Shannon,Zhu1992-general-Nyquist,Luke1999-origin-sample-thrm} proves that the maximum step size between scan points  of $\sfrac{\lambda}{2}$ is required for a complete description of the beam.  %If the beam pattern measurement is taken at a second focal plane with a magnification factor $M$, then the sampling criterion changes to $\sfrac{M\lambda}{2}$.  
However, scans at smaller step sizes (oversampling) can extract more information from the measurement, for example reflections and beam steering \cite{Weisstein2014-mathematica-nyquist}. Step sizes as small as $\sfrac{\lambda}{10}$ or smaller are not uncommon for single pixel measurements.  For large focal planes of many pixels there is a trade-off between number of sampling points across the field of view and system stability over long scan durations. 

\section{The Gaussian-Hermite Field Expression} \label{sec:gauss-beam-eqs}

In the THz frequency regime, the wavelength of the received radiation is comparable to the dimensions of the optical elements it interacts with, the telescope beam will suffer diffraction effects such that ray tracing is no longer valid as it is in the optical regime. Instead, the beam is better modeled as a Gaussian function, and so instrument scientists often choose to design their optical systems using Gaussian optics principles. This is also a matter of convenience, because in Fourier optics a Gaussian beam transforms to another Gaussian beam, allowing a single set of equations to describe the beam at all points in the optical system. The diagnostic power of a beam pattern measurement thus comes from the ability to measure the field pattern of an instrument and fit it to a fundamental field function so that the instrument's performance can be modeled and verified. 

This section gives an overview and mathematical description of an idealized Gaussian beam, and section \ref{sec:beamfitting-tech} will discuss a method to take a measurement of $S(x,y,z)$ and fit it to this function (here $S$ is a generic term for instrument response).  Other field functions can describe the nature of beams in this frequency regime, for example a truncated Bessel function \cite{2012-BesselFunctions,Yousif1997-bessel-cpx}, but are not considered in this analysis. 

A Gaussian function, in terms of a 2D power function $P$, has a centralized peak $P_o$ at a centroid location $x_o$,  and falls off exponentially and symmetrically as a function of distance $\pm x$ from the centroid location. Therefore, $P(x_o) = P_o$ and 
\begin{equation}
P(x) = P_o~exp\left(-2\frac{x^2}{\omega(z)^2}\right)
\label{eq:gauss_power}
\end{equation}
where $\omega$ is the width of the function measured from the central axis at $z$, defined as the point where the power amplitude drops to $\sfrac{1}{e}$ of the on-axis strength (so where $P(x) = \sfrac{P_o}{e}$). Here we have adopted an $(x,z)$ planar coordinate system. The term $\omega$ is referred to as the beam radius, since it is measured from the axis defined by $x_o$. A convenient reference point for a Gaussian function is the point $z$ at which the function of $P(x)$ is dropped by a factor of two. We refer to this point as the full width at half maximum (FWHM) point. In dB scale, this value occurs at the -3 dB point of the beam. The relationship between the FWHM and the beam radius is $\mathrm{FWHM} = 2 \sqrt{2\ln(2)}~\omega(z)$.  

In an optical system, we now consider not a 2D Gaussian power function but a 3D propagating electromagnetic field, so  $P\rightarrow \vec{E}$, where $\vec{E}$ describes a sensitivity pattern (the beam).  The field $\vec{E}$ is complex, meaning it has both real and imaginary components. At any field point $\vec{E}(x,y,z) = \alpha \pm i\beta$, the amplitude $A$ of the beam is  $A=\sqrt{\alpha^2 + \beta^2}$, and the phase $\phi$ of the beam is $\phi=\arctan{\left(\sfrac{\beta}{\alpha}\right)}$.

\begin{figure}
	\centering
	\includegraphics[width=5.5in]{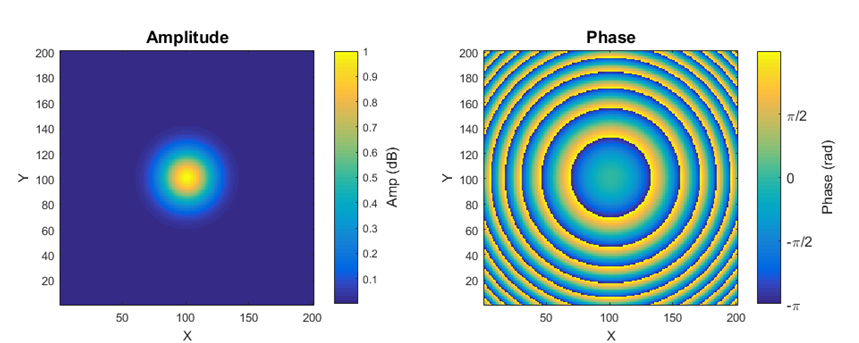}
    \caption[~Idealized Gaussian Beam-Amplitude, Phase, and 3D Pattern]
    {The upper two panels show the amplitude (left) and phase (right) of an
    idealized fundamental Gaussian beam produced using equation
    \eqref{eq:gauss.hermite},
    both as viewed in a 2D plane 
    perpendicular to the optical axis (i.e. from $z_{ref}$). 
    }
 \label{fig:2D-3D-beams}
\end{figure}

The two panels of figure \ref{fig:2D-3D-beams} shows an ideal Gaussian beam at the reference plane $z_{ref}$ as viewed from a point further on the optical axis. The left panel shows the amplitude pattern viewed form a plane and the right panel shows the phase pattern. In the ideal case the beams are circularly symmetric along the central optical axis. We see the phase peak at this point with a symmetrical (spherical) roll-off as a function of distance $r = \sqrt{x^2+y^2}$ from the axis. The plane shows concentric rings, where each ring shows a phase ‘jump’ from $\sfrac{-\pi}{2}$ to $\sfrac{+\pi}{2}$.

\begin{figure} 
	\centering
	\includegraphics[width = 4.5in]{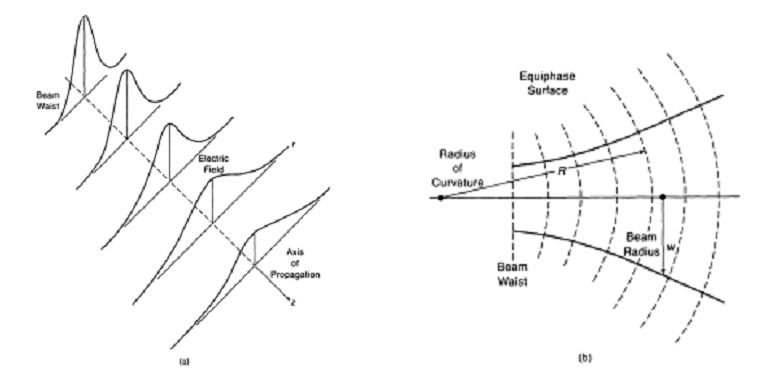}
    \caption[~Gaussian Beam Propagation Example]
    {Left panel shows a 1D Gaussian beam as it propagates along the $z$ axis. 
    The beam at the focal plane $z_0$ initially has a beam waist 
    $\omega_0$. As the beam travels along the $z$-axis, the beam radius $\omega(z)$
    gets larger, and the peak amplitude decreases. In the right panel, we look
    at the beam along the yz plane. The solid lines represent the amplitude of the
    beam, and the dashed lines show the phase front of the beam. 
    The phase radius $R$ is clearly
   referenced to a point behind the focal plane.  
   This figure is reproduced from Goldsmith 1998
  \cite{GoldsmithPaulF1998-GaussBeamsTxtbook}, 
  with permission.}
  \label{fig:goldsmith}
\end{figure}

Gaussian beams do not converge into a single focal point, but rather come to a narrowest convergence point at the focal plane. The radius of the beam at that location is called the beam waist $\omega_0$,  indicated in panel b of figure \ref{fig:goldsmith}. As the beam propagates outward from this position, the amplitude peak of the Gaussian beam spreads out, getting wider and less prominent. Therefore the beam radius is a function of the distance $z$ from the focal point, $\omega(z)$. Panel b of figure \ref{fig:goldsmith} also shows the phase of the emerging beam is constant over a spherical surface of a radius $R$. The location of the phase center is \emph{not} the location of the beam the  waist $\omega_0$. If we call the beam waist location $z = 0 \Rightarrow z_0$ then the phase center is located at negative $z$, or $z<0$. %However, as $z \to \infty$,  in equation \eqref{eq:GBradius} we see that $R \to z$, and thus the phase center approaches $z_o$ for large values of $z$. 

The full mathematical field function of a propagating Gaussian beam in a 3D Cartesian coordinate system can be described by the first-order Gaussian Hermite polynomial, given in the equation \eqref{eq:gauss.hermite}. Here, we ignore the time dependence of the equation, which is required for beam propagation but is not measured with a radiation pattern measurement. A full discussion of Gaussian beam propagation is the subject of Goldsmith 1998 \mbox{\cite{GoldsmithPaulF1998-GaussBeamsTxtbook}}. 

\begin{equation} \label{eq:gauss.hermite}
	\vec{E}(x,y,z) = 
	\left( \dfrac{2}{\pi \omega_x \omega_y} \right) ^ {\frac{1}{2}}
	\exp{\left[ 
    \left(-\dfrac{x^2}{{\omega_x}^2} - \dfrac{y^2}{{\omega_y}^2}\right)
	-1i \left(\dfrac{\pi x^2}{\lambda R_x} + \dfrac{\pi y^2}{\lambda R_y}
	- \dfrac{\phi_{0,x}}{2} - \dfrac{\phi_{0,y}}{2}\right)
	\right]}
\end{equation}

The terms $\omega_x, \omega_y, R_x, R_y, \phi_x$, and $\phi_y$ can all be re-written in terms of $\omega_o$ and $z$ using the following equations:
\begin{equation} \label{eq:wz}
	\omega_n(z) = \omega_{0,n}
	\left[ 1+
    {\left( \dfrac{\lambda z}{\pi {\omega_{0,n}}^2}
    \right)}^2 \right] ^{\frac{1}{2}}
\end{equation}
\begin{equation} \label{eq:GBradius}
	R_n(z) = z+ \dfrac{1}{z}
	{\left( \dfrac{\pi{\omega_{0,n}}^2}{\lambda}
    \right)}^2
\end{equation}
\begin{equation} \label{eq:GBphase}
	\phi_{0,n}(z) = 
    \tan^{-1}{ \left(
    \dfrac{\lambda z}{\pi {\omega_{0,n}}^2}
    \right) }
\end{equation}

where the subscript $n$ can indicate either the $x$ or $y$ axis. Therefore, a Gaussian beam observed in an arbitrary measurement plane can be described by the fundamental parameters $\omega_{0,n}$ and $z$.

Breaking down equation \eqref{eq:gauss.hermite}, we see that that the fundamental parameters can be separated into a scalar term and an exponential term. The scalar term is a normalizing factor such that the integrated power in the beam adds to unity. The exponential term can be broken down further into real and imaginary terms. The real parts of the exponential term describe the amplitude behavior of the beam, and the imaginary parts of the exponential term describe the phase behavior of the beam. Looking first at the real (amplitude) terms of exponential in equation \eqref{eq:gauss.hermite}, we see that the first term determines the shape of the Gaussian beam.

\begin{figure}
	\centering
	\includegraphics[width=2.5in]{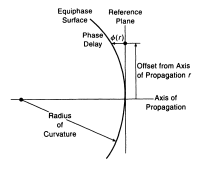}
    \caption[~Phase Behavior of Gaussian Beams]
    {Representation of the phase terms of equation
    \eqref{eq:gauss.hermite}
    as referenced to a plane at $z_{ref}$. 
    Figure reproduced from Goldsmith 
    1998 \protect{\cite{GoldsmithPaulF1998-GaussBeamsTxtbook}},
    with permission.}
    \label{fig:gauss-phaseterms}
\end{figure}

Looking towards the phase (imaginary) exponential terms, the spherical phase fronts are described by the phase front $R$, which measures the the distance between the phase center and the reference plane at $z_{ref}$ along the optical axis. The terms $R_x(z)$ and $R_y(z)$ describe the phase radius of each component of the wave in the $xz$ and $yz$ planes, respectively. Note that the phase surface in the plane at $z_{ref}$ is only equal to $R_{x}$ at a single point, which is the center of the plane at $(x=0,y=0,z_{ref})$. Elsewhere in the plane $(x\neq0,y\neq0,z_{ref})$, the phase surface is delayed by the factor $\phi_x(x,z)$ and $\phi_y(y,z)$, as represented in figure \ref{fig:gauss-phaseterms}. 

In \eqref{eq:gauss.hermite}, we see that that the fundamental parameters are independent along the $\hat{x}$ and $\hat{y}$ axis. This allows us to fit for beam asymmetries. The beam waist $\omega_0$ can differ along the $\hat{x}$ and $\hat{y}$ axis, so $\omega_{0,x} \neq \omega_{0,y}$. The ratio of $\sfrac{\omega_{0,x}}{\omega_{0,y}}$  is a measure of the beam’s ellipticity. Phase centers that are not aligned along the optical axis are described as astigmatic beams. Note here that a beam ellipticity implies astigmatism, though the two terms refer to different characteristics of the beam. The difference in distance between the phase centers of the beam is measured by the term $\delta z_{x,y}$, using the notation presented in \mbox{\cite{Jellema-PhD2015}}. The $z$ terms in equations \eqref{eq:wz}-\eqref{eq:GBphase} can be substituted for $z \pm \delta z_{x,y}$ for one axis in the case of astigmatic beams. The idealized case of figure \ref{fig:2D-3D-beams} occurs when $R_x = R_y \Rightarrow R_{x,y}$ and $\phi_x = \phi_y \Rightarrow \phi_{x,y}$. 

For Gaussian beams, there is a natural distinctive boundary between two regions closer and further from the beam. As seen in the amplitude (solid lines) in the lower panel of figure \ref{fig:goldsmith}, close to the beam waist there is a region where the amplitude does not vary significantly as a function of distance, and the beam is roughly collimated. However, in that region, the phase of the beam is changing rapidly with $z$, beginning as planar at the beam waist and becoming more curved as the beam travels in $z$. At some distance, the phase reaches its maximum curvature, after which the propagating beam spreads out and approaches a flat phase front. The location of the minimum radius of curvature is called the confocal distance, given by the equation 
\begin{equation} 
	z_c = \dfrac{\pi {\omega_0}^2}{\lambda}
    \label{eq:confocal}
\end{equation}
where $z_c$ is the confocal distance. 

In approximate terms, the near-field region of the beam is where the beam is roughly collimated in amplitude but varying widely in phase as a function of $z$, and the far-field region is where the beams begin to be consistent, well-defined, and approximated as plane waves. The distinction becomes important for radiation pattern measurements because the beam can behave very differently when measured in each region. However, there are advantages to measuring the beam in either the near field or far field as discussed in section \ref{sec:introNF2FF}.

There is no precise cutoff distance that distinguishes the near and far fields, but the confocal distance of Gaussian beams agrees with generally agreed-upon values (see section 2.2.4 of \cite{GoldsmithPaulF1998-GaussBeamsTxtbook}). A more rigorous explanation involves the reactive near-field, where the presence of evanescent modes dominates. Evanescent modes are solutions to the field function that do not propagate, and decay exponentially as a function of $z$. However, there is a region very near the radiating element (the reactive near field, $z \lessapprox \lambda$) where the signal contained in these fields is still significant enough that these modes affect the beam pattern. 

Further into the near field, sometimes referred to as the Fresnel region, the beam is still diffracting with itself, causing rapid change in the waveform shape as a function of distance. This region is so named after the Huygens-Fresnel Principle, which states that each point on an arbitrary waveform is a secondary source of its own spherical waveform. To find the shape of a new, secondary waveform at some forward distance, the contribution of all points at the primary waveform must be summed for each point on the secondary waveform, including amplitude decay and phase delay of each contributing wave along the propagation distance to the point on the secondary waveform. The diffraction caused by the interaction between all of the waveforms leads to significant change in the beam pattern as a function of distance. After some distance, these effects become negligible, and the far-field radiation pattern dominates. 

\section{Analysis Techniques Using Complex Field Measurements } \label{sec:analysis-techs}

A significant advantage to complex versus direct field measurements is that the intensity, beamwidth, and pointing direction are measured simultaneously, so only a single scan plane is required to determine basic beam characteristics. A single complex field pattern can be propagated and recreated at any plane along the optical axis, and so the scan can be conducted at arbitrary planes in the optical path, including in the near-field. This way, the beam pointing direction can be found by propagating the beam mathematically, rather than by measuring it in multiple separate planes as with intensity measurements. Complex or compact optical systems may only have one plane accessible to a beam scanning system, and thus being able to characterize the beam from a single scan plane is ideal. 

Limitations in source probe output power at THz frequencies make it very difficult to achieve full end-to-end optical system characterization including a telescope's primary and secondary mirror. For example, let us use the definition of the far-field boundary in \eqref{eq:confocal} but replace $\omega_o$ with $D$, the diameter of a telescope's primary mirror. For a telescope with a 1-meter primary operating at a wavelength of $\lambda=$ 1mm, the far-field boundary occurs at 1 km away from the telescope aperture. Typical output power of source probes in this regime are not adequate to overcome the atmospheric attenuation at such long distances, making this measurement impractical from a technological as well as experimental configuration standpoint. 

Though a single complex beam scan can characterize an instrument's sensitivity pattern, pointing direction, and coupling efficiency, higher-level information can be extracted for full beam characterization by conducting two sets of measurements at two scan planes using a singly polarized source probe. Two sets of scans are conducted each at distance $z_1$ and $z_2 = z_1 \pm \frac{\lambda}{4}$ , where the choice of $z$ can be located either in the near- or far-field of the instrument.  The quarter-wave offset in distance can be used to correct for optical standing waves introduced between the between the source probe and the optical system (i.e. standing waves between the source probe and cryostat). At each plane, after the first measurement the source probe is rotated by 90$^\circ$ and the beam pattern is re-measured, which can later be used to fit for the co- and cross-polarization axis of a receiver beam. Each of these techniques will be discussed in detail below.  

\subsection{Standing Wave Reduction} \label{sec:qw-intro}

The principle used in this analysis to remove standing waves can, to first order, eliminate the effects of optical standing waves throughout the optical system by virtue of the principle of linear superposition. In post-processing of the data from the two scan planes $z_1$ and $z_2$, we effectively cancel the standing waves using the equation \ref{eq:QWreduction}

\begin{equation}
	\vec{E}_{comp} = \dfrac{\vec{E}(z_1) + \vec{E}(z_2)e^{\pm \frac{i\pi}{2}}}{2}
    \label{eq:QWreduction}
\end{equation}

where $\vec{E}(z_1)$, $\vec{E}(z_2)$ are the complex fields measured at each scan plane, and $\vec{E}_{comp}$ is the compensated field. The exponential term will be negative for $z_2$ further from the receiver than $z_1$, and positive otherwise.  When the two maps are co-added, a wave traveling parallel to the optical axis will have a phase shift of $\frac{\pi}{2}$, but a standing wave, traveling twice the distance, will have a phase shift of $\pi$. These waves will cancel while the primary beam is simply averaged together. $\vec{E}_{comp}$ can be used for further post-processing. A more detailed description of this technique can be found in Jellema 2015 \cite{Jellema-PhD2015}.  

\subsection{Polarization} \label{sec:pol-methods}

\begin{figure}
   \centering
   \includegraphics[width=3in]{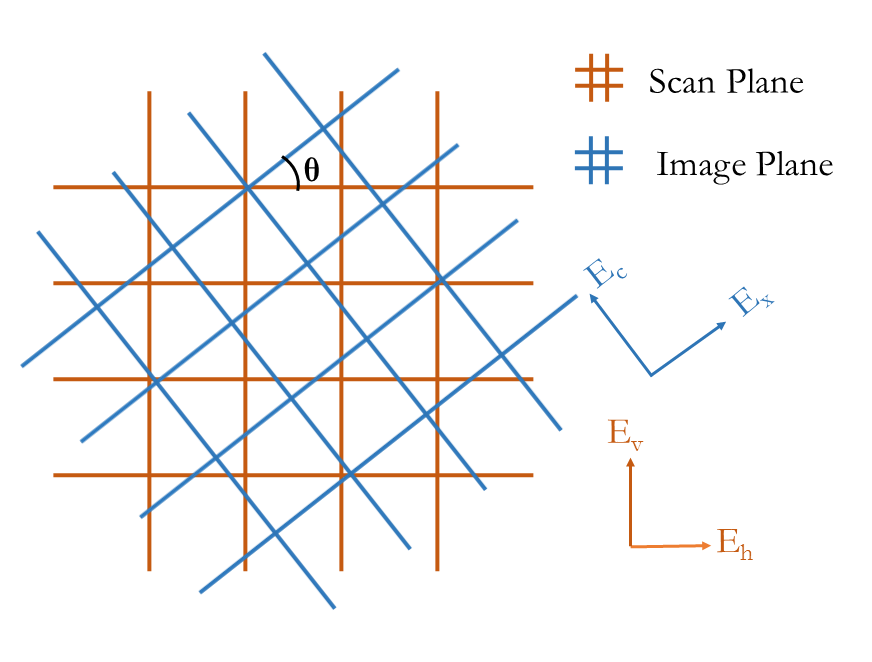}
      \caption[~Example of Rotating Polarized Scan Coordinates]
      {Example of a scan plane recorded at an arbitrary polarization
      orientation with respect to the co- and cross-polarization 
      orientation of the instrument. A complex field measurement
      allows the measured scan plane to be rotated mathematically, 
      and a fitting function can rotate the scan plane until
      the signal in the cross-polarization field is minimized. }
   \label{fig:grid}
\end{figure}

% Co and Cross-polarization 
Polarization orientation is a key figure of merit for astronomical instruments, whether they be singular or dual-polarization sensitive. The co- and cross-polarized field components can be measured if the polarization axis are known or measured, which are the field components that can be compared directly to optical simulations. In order to measure the polarization orientation of an instrument, the complex field pattern can be measured with either a dual-polarization source probe of which the polarized field pattern is known and corrected for, or it can be measured with two scans of a singly polarized source.

In the latter case, the source probe must be rotated by precisely 90$^\circ$ between the two scans. If the source probe orientation is precisely aligned with the co-and cross-polarization axes, the two scans measure the co- and cross-polarization fields. However, if the instrument's principle axes are unknown or need to be measured, complex field measurements taken at an arbitrary scan coordinate system can be transformed to other axes, and the co- and cross-polar orientation can be found by maximizing or minimizing the total integrated power in the field, respectively.  For array instruments, this has the advantage of fitting for a polarization orientation individually for each pixel. 

In the case of a singly-polarized source probe, to almost identical scans are completed, changing only the source probe orientation. The probe can initially be aligned arbitrarily, and rotated for the second scan. Here we designate the two complex fields as $\vec{E}_h$ and $\vec{E}_v$. These measurements can be transformed onto new axes $\vec{E}_c$ (co-polar) and $\vec{E}_x$ (cross-polar) by a simple matrix rotation of the two fields by a rotation angle  $\theta$, which is give in equation \eqref{eq:pol_mtx}.  A minimization algorithm can be applied to solve for the transformation that minimizes power in the cross-polar field $|\vec{E}_x|$.  Figure \mbox{\ref{fig:grid}} shows a simple illustration of this transformation.  
. 
\begin{equation}
  \left[ \begin{array}{c}
      \vec{E}_c\\
      \vec{E}_x
  \end{array} \right]
  =
  \left[ \begin{array}{cc}
  		\cos(\theta)&\sin(\theta)\\
        -\sin(\theta)&\cos(\theta)\\
   \end{array} \right]
   \left[ \begin{array}{c}
  		\vec{E}_h\\
        Ae^{i\phi}\vec{E}_v\\
   \end{array} \right]
   \label{eq:pol_mtx}
\end{equation}

 The term $Ae^{i\phi}$ is a scaling factor in amplitude $A$ and phase $\phi$. The scaling factor corrects for system drifts between the two measurements $\vec{E}_{h}$ and $\vec{E}_{v}$. This term also corrects for the amplitude factor between the two maps to account for the difference in coupling between one axis $h$ or $v$ that is more aligned to the co-polar axis during the scan.

Equation \eqref{eq:pol_mtx} can be used within a nested algorithm to calculate $\vec{E}_{c}$ and $\vec{E}_{x}$. Equation \eqref{eq:pol_mtx}  is iterated over a set range of rotation angles $\theta$  . For each iteration, a Nelder-Mead \cite{Nelder1965} minimization function finds the optimal scaling parameters $A$ and $\phi$. Each iteration is initialized with an initial guess of $A$ and $\phi$ taken from the center of each measurement, such that $A_{guess} = \operatorname{Re}[\vec{E}_h(x=y=0)] - \operatorname{Re}[\vec{E}_v(x=y=0)]$ and $\phi_{guess} = \operatorname{Im}[\vec{E}_h(x=y=0)] - \operatorname{Im}[\vec{E}_v(x=y=0)] $.  The function selects the angle $\theta$ with the lowest value of $|\vec{E}_{x}|$, using the values of $A$ and $\phi$ found with the minimization algorithm. 

\subsection{Beamfitting} \label{sec:beamfitting-tech}

Section \ref{sec:gauss-beam-eqs} presented an overview of Gaussian beam formalism, and this section will describe the technique to fit a radiation pattern measurement to equation \eqref{eq:gauss.hermite}. This is achieved by calculating the degree of coupling between the measured complex field $\vec{E}_{m}$ (possibly after processing detailed above) to an idealized beam $\vec{E}_{idl}$ by equation \eqref{eq:overlap_int}

\begin{equation}
	c_{00} =                                          
    \dfrac
    {\iint
   \vec{E}_{idl}\vec{E}_{m}^* dxdy}
    {\iint \sqrt{{|\vec{E}_{idl}}|^2}dxdy
     ~\iint \sqrt{{|\vec{E}_{m}}|^2}dxdy}
     \label{eq:overlap_int}    
\end{equation}
where $\vec{E}_{ideal}$ is the result of equation \eqref{eq:gauss.hermite}. The beamfitting algorithm initiates $\vec{E}_{idl}$ using the designed of parameters $\omega_{o,x}, \omega_{o,y}$ at the nearest focal plane and propagates the idealized beam forward to $z_m$, the measurement plane. Gaussicity is the maximum fraction of coupled power into a fundamental Gaussian beam which best approximates the measured field. Gaussicity $\eta$ can be calculated from the coupling parameter by  $\eta = |c_{00}|^2$. The coupling loss between the measured beam and the idealized beam is this value subtracted from unity, or $\epsilon = 1- \eta = 1-|c_{00}|^2$.

\begin{figure}
	   \centering
	\includegraphics[width=\hsize]{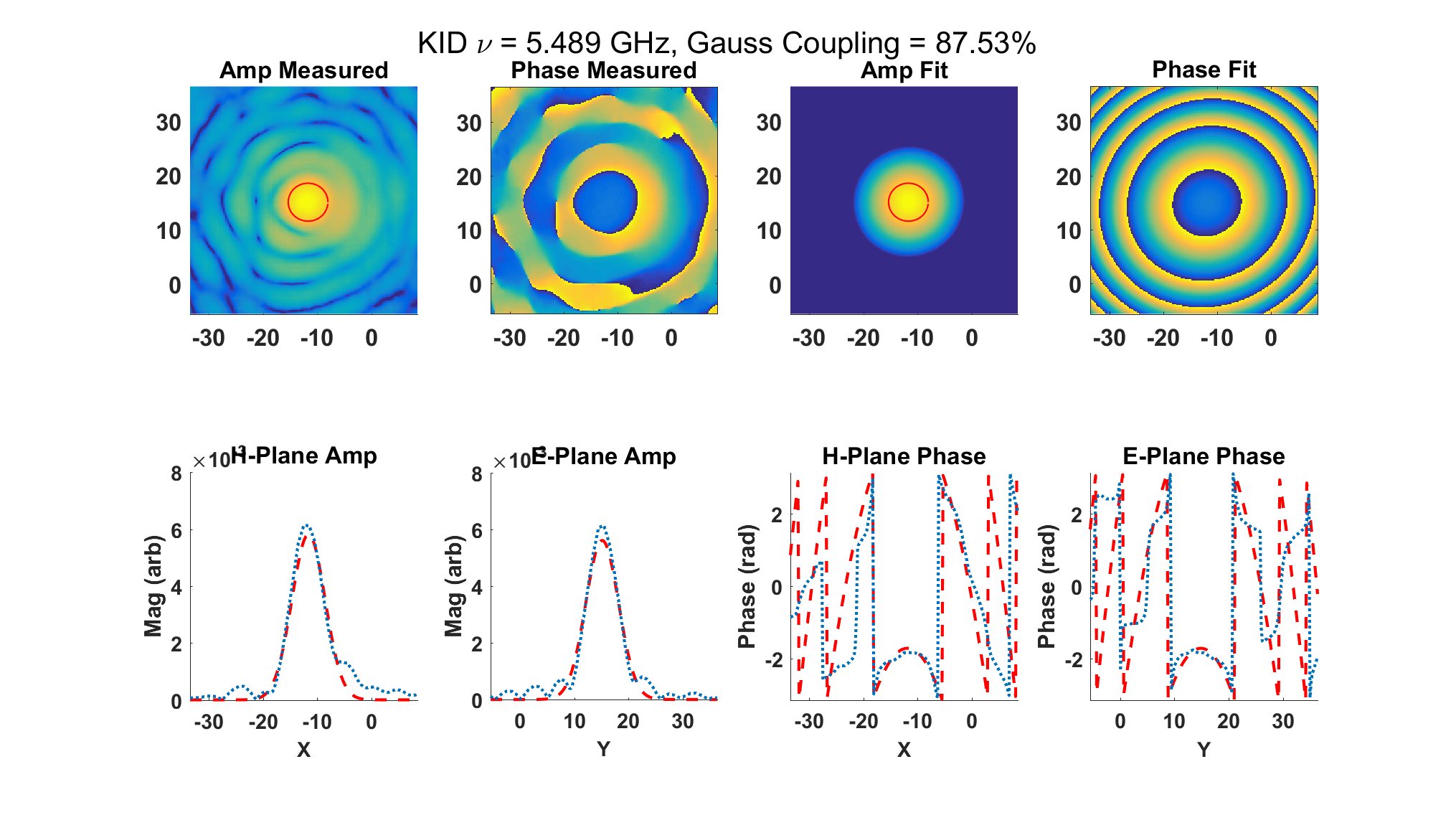}
    \caption[~Beamfitting to Measurement 2D Data Example]
    {Comparison between a measured near-field beam
    pattern and the Gaussian beam
    $\vec{E}_{idl}$ produced by equation 
    \eqref{eq:gauss.hermite}. The upper panels show the 2D 
    data and the lower panels show 1D cuts in the E- and 
    H-plane. The blue dotted lines 
    are the measured data and the dashed red 
    lines are the fit data. The red circle in the
    upper two amplitude panels indicates the FWHM beam width.}
  \label{fig:beamfit-comparison}
 \end{figure}

If instead we do not assume the beam parameters $\omega_{o,x}, \omega_{o,y}$ and want to fit for the values $\omega_{x}(z), \omega_{y}(z)$, we can search the parameter space using a minimization function operating on the coupling loss parameter $\epsilon$. The minimization function uses 'seed' values as initial guesses of the beam parameters, computes the value of $\vec{E}_{idl}$ from these values, propagates the beam forward to the image plane, calculates the coupling loss coefficient between the measurement and fit data,  and iterates over the parameter space until a convergence criteria is met. Figure \ref{fig:beamfit-comparison} shows a comparison between an example of a  measured beam pattern and the best-fit beam produced with this technique.  

The minimization function can be specified depending on the degree of freedom given to the search criteria. An unbounded Nelder-Mead~\mbox{\cite{Nelder1965}} minimization function can efficiently probe the parameter space and has a low chance of getting stuck into local rather than global minima. As an added processing step, the output parameters of that function can be used as an initial guess for a non-linear least-squares minimization function~\mbox{\cite{Marquardt1963-lsq-algorithm}}, which is also unbounded. The least squares algorithm can more easily find the confidence intervals for the solution set of beam parameters. 

\begin{figure}
	\centering
	\includegraphics[width =4.5in]{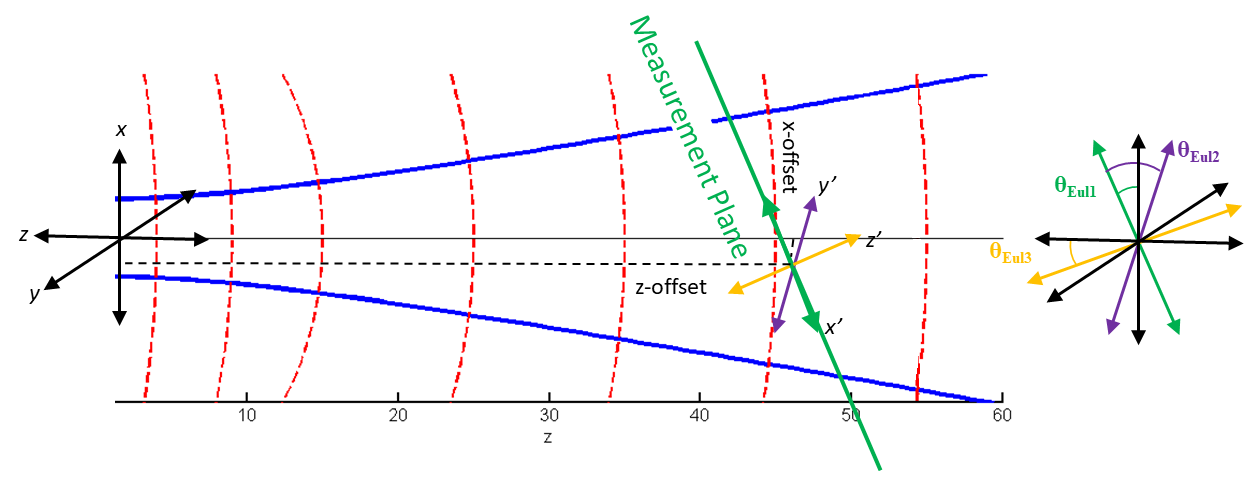}
    \caption[~~ Scan Plane to Image Plane Coordinate Transformation Diagram]
    {Demonstration of a Gaussian beam in a reference
    frame $x,y,z$, and a measurement plane with a misaligned 
    coordinate system $x’,y’,z’$. The idealized Gaussian beam 
    parameters must be transformed from the $x’,y’, z’$ to the 
    $x,y,z$ system before the overlap integral can be performed. 
    The blue lines show the beam amplitude, and red dashed lines
    show the spherical phase fronts. The primed and unprimed 
    coordinate systems are shown in relation to their origin 
    and are also superimposed to the right of the beam diagram.}
 \label{fig:pc-coords-gaussbeam}
\end{figure}

It is important to note that a radiation pattern is measured relative to the coordinate system set by the measurement plane, which may not be along the principle axis of the receiver if there are lateral or rotational offsets between them. These offsets will skew the measurement plane relative to the image plane and can cause a beam to appear astigmatic or asymmetrical if not properly accounted for. One way to correct for this is to very precisely align the coordinate system of the scan plane to the optical axis of the receiver. Doing so requires high precision metrology of the apparatus prior to the measurement. However, as was shown with Jellema 2015~\mbox{\cite{Jellema-PhD2015}}, the frequency dependence of the optical behavior of a system can cause apparent misalignments in a beam scanning system, even if properly aligned using laser metrology. 

A better course of action is to mathematically fit for the lateral and rotational offsets of the scanner system with respect to the optical plane. In the beam fitting algorithm, an idealized beam is initiated in an arbitrary coordinate system with lateral offsets $x_{off},y_{off},z_{off}$ and rotated  by $\theta_{Eul1}, \theta_{Eul2}, \theta_{Eul3}$ with respect to the scan plane. This new coordinate system $x',y',z'$ is used to propagate an idealized, fundamental Gaussian beam $\psi_{00}$.  The angles $\theta_{Eul1}, \theta_{Eul2}, \theta_{Eul3}$ are Euler-rotation angles~\mbox{\cite{EulerRotation-Robotics2012}}. Figure~\ref{fig:pc-coords-gaussbeam} shows the coordinate system transformation described here. 

\subsection{ Beam Propagation}
\label{sec:beam-prop}

\begin{figure}
\centering
	\includegraphics[width = \hsize]{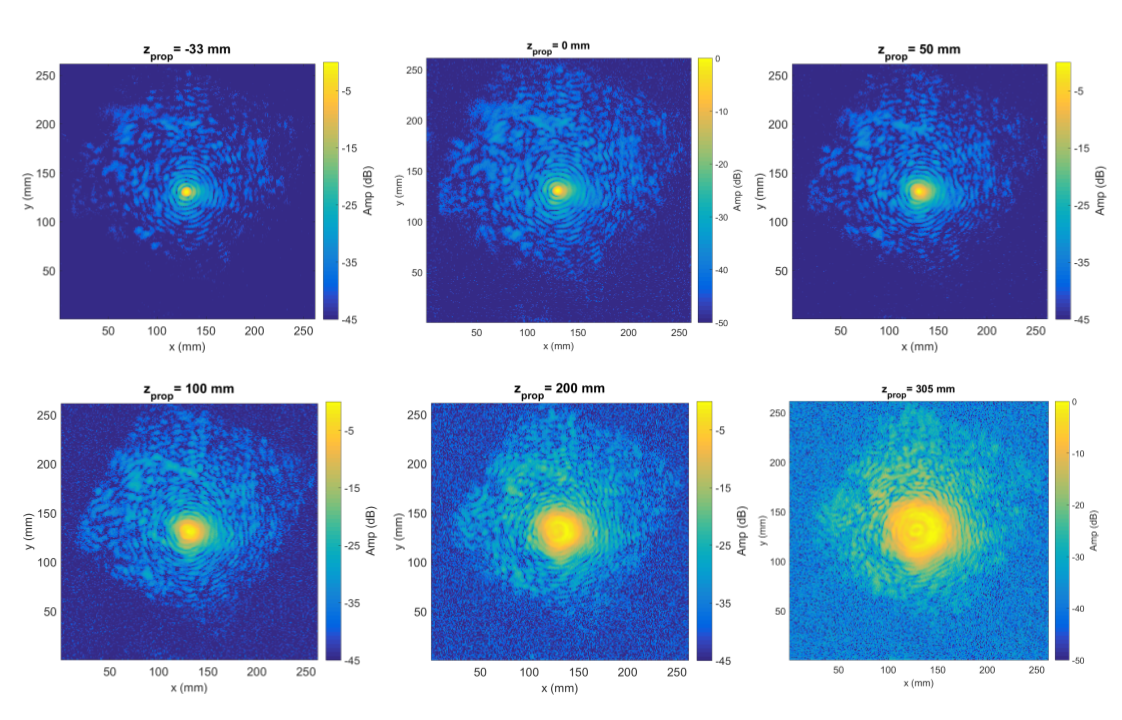}
	\caption[~Beam Propagation]
    {Demonstration of propagating a beam from the measurement plane 
    backwards to the focal plane (here -33 mm) and outward towards
    the laboratory environment. The choice of propagation distance 
    is arbitrary and is used here to demonstrate the beam's 
    diversion. In this demonstration we recreate the beam from 
    unmasked near-field data.}
    \label{fig:beam_prop}
\end{figure}

One advantage of complex field radiation pattern measurements is the ability to propagate the measurement plane either forwards or backwards through optical elements to arbitrary planes along the principle axis. There are two common methods used to propagate fields, though both are approximations. The first is based on a Modal Expansion technique as described in Balanis 2005 \cite{BalanisConstantineA.2005-AntennaTheory}. The second technique is based on the angular plane wave spectrum representation (APWS) \cite{APWS-lecturenotes-online,Tervo2002-APWS-coherentEMfields,Hollis1973-NFmath}, and will be described here. Though the APWS method of beam propagation is a highly accurate and representative, it is still only an approximation, in the same way a Fourier transform approximates an original function. 

The APWS technique is analogous to a 2D Fourier transform converting a timeseries measurement to a frequency. In a 1D Fourier transform, a function is represented by an infinite series of sine (or cosine) functions. The sine functions each have an independent magnitude, frequency and phase offset, and the superposition of the infinite series of sine and cosine functions approaches a perfect description of the original function. For the 2D case, rather than breaking up a function into a infinite series of sines and cosines, the function is broken down into an infinite series of plane waves. The series of plane waves is the Fourier transform of the near-field measurement $\vec{E}_{nf}$, and the series is represented by the variable $\vec{A}$, after the Angular Plane Wave spectrum method (though in some sources it may be referred to as $\hat{E}$, as is common to designate a Fourier pair).  

We perform a 2D FFT of the complex field $\vec{E}_{nf}$ measured in the near-field to find the APWS using the equation 
\begin{equation} \label{eq:measplane2apws}
	\vec{A}(k_x,k_y) = 
    \frac{1}{({2\pi})^2}
    \int_{-\infty}^{\infty} \int_{-\infty}^{\infty}
    \vec{E}_{nf}(x,y,z=0)
    e^{-i(k_xx+k_yy)}
    dxdy
\end{equation}
The field $\vec{E}_{nf}$ contains all the information about how a beam's sensitivity, pointed along the optical axis $\hat{z}$, varies as a function of spatial coordinates across the plane. In the reciprocal space, the Fourier field $\vec{A}$ contains the information of the beam's sensitivity as a function of angle, and all of the plane waves are co-aligned at the origin. From an optics standpoint, the field $\vec{A}$ represents the pupil plane (aperture) of an optical system, where each point in the object plane fills the aperture, but has an individual pointing direction. 

In the spectrum $\vec{A}$, all of the plane waves share the same wavelength (frequency), but each wave is pointed in a different direction and has a different vector length, proportional to the magnitude of the field in that direction. An individual wave is described by its propagation vector $\vec{k}$, which is pointed in the direction of propagation, which is normal to the plane, and has units of $\mathrm{\sfrac{1}{m}}$. The individual propagation vector can be projected onto a set coordinate system, so the pointing direction of each plane wave can be referenced to the original coordinate axis. We can then describe the propagation vector using $\vec{k} = k_x \hat{x}+k_y \hat{y}+k_z\hat{z}$, where the angle defined by $\tan({\sfrac{k_y}{k_x}})$ describes the pointing angle of the beam, and $k_z$ represents the propagation along the $\hat{z}$ axis. 

%Since we do not yet consider the variation of the wave along the $\hat{z}$ direction in equation \eqref{eq:measplane2apws}, we only include $k_x,k_y$ terms in the angular plane wave spectrum $\vec{A}$. 
The angular resolution (the wavenumber spectrum points) of the plane waves making up $\vec{A}$  is dependent on the sampling of the $\vec{E}$ field in the measurement plane. From a sampling of the complex field over the $\hat{x}$ and $\hat{y}$ coordinate system at regular grid spacing of $dx$ and $dy$, we can construct a grid of $M\times N$ points such that $-\frac{M}{2} \leq m \leq \frac{M}{2}-1$ and $-\frac{N}{2} \leq n \leq \frac{N}{2}-1$.  Nyquist sampling theorem places an upper limit on the spacing of the grid points $dx,dy$ of $\sfrac{\lambda}{2}$ in order to properly reconstruct the field. The extent of the sampling plane $M,N$ is typically reaches at least -30 dB from the amplitude maximum of the beam. 

The values of the wavenumber spectrum points (angles) on the grid in Fourier space are: 
\begin{eqnarray}
	k_x = \dfrac{2\pi m}{Mdx}\\
    k_y = \dfrac{2\pi n}{Ndy}
\end{eqnarray}

\begin{figure}
	\centering
    \includegraphics[width=5in]{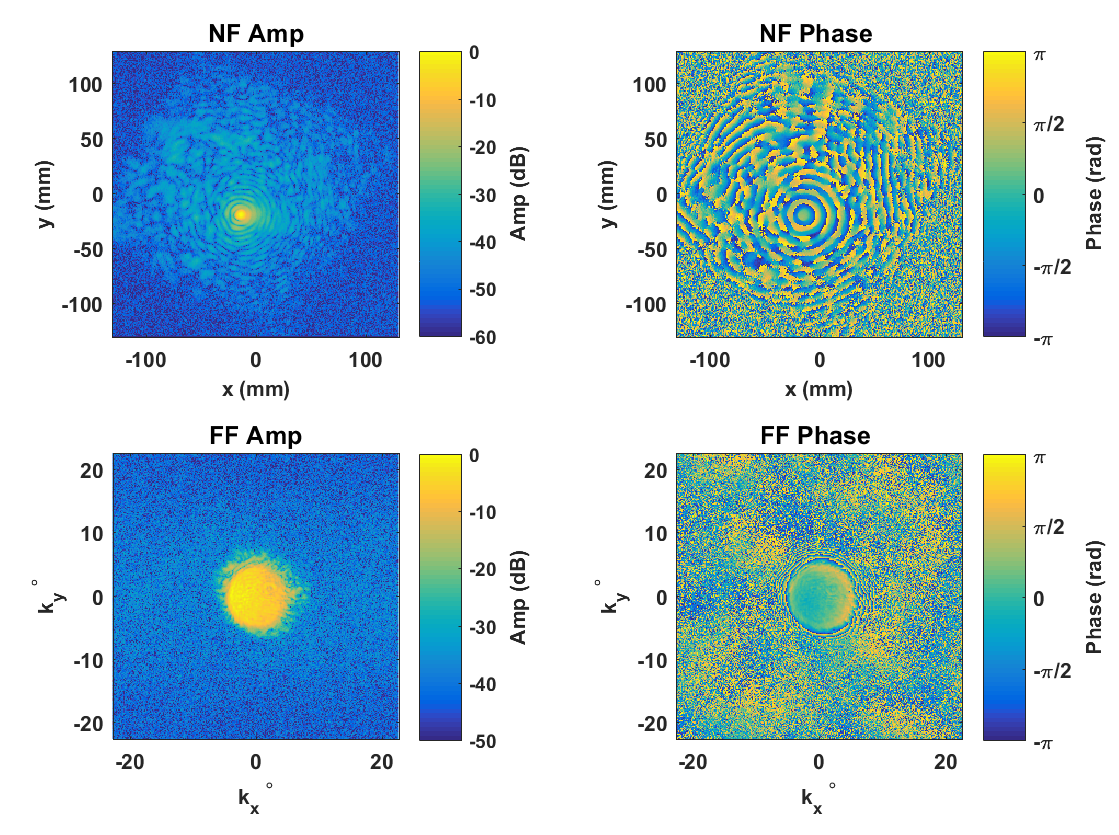}
    \caption[~~ Near-Field to Far-Field Propagation]
    {The upper two panels show the amplitude (left) 
    and phase (right) of $\vec{E}_{nf}$ and the lower two 
    panels show the transformed field $\vec{A} = \vec{E}_{ff}$. Data for 
    this figure represents a single pixel of a wide-field MKID detector array.}
	\label{fig:NF2FF}
\end{figure}

In some references, because of the analogy to 1D Fourier transforms between time domain sampling and frequency, the coordinates are referred to as 'spatial frequencies', though this can lead to some confusion because we ignore temporal dependence in the plane wave representation. So when viewing the field $\vec{A}(k_x,k_y)$ in a 2D map, the value of $|\vec{A}|$ is proportional to the intensity of the wave at each grid point ($k_x,k_y$), where each grid point represents the angle or pointing direction $\tan({\sfrac{k_y}{k_x}})$ across the 2D plane. 

So far we have only discussed the plane wave spectrum as a stationary field, located at the measurement plane. We now consider the $z$ dependence of the spectrum. This is represented by the $k_z\hat{z}$ term in the propagation vector $\vec{k}$. In order to propagate the plane wave spectra to an arbitrary plane located at $z\neq0$, the field must be multiplied by a propagation factor $e^{\pm ik_zz}$ such that
\begin{equation}
\vec{A}(k_x,k_y,z) = \vec{A}(k_x,k_y)e^{\pm ik_zz}
\end{equation}
where
\begin{equation} \label{eq:kz}
	k_z = \pm \sqrt{{k_o}^2-{k_x}^2-{k_y}^2}
\end{equation}
and
\begin{equation} \label{eq:ko}
	k_o = \dfrac{2\pi}{\lambda}
\end{equation}

In order to satisfy the Helmhotz equation, the two solutions for $\pm k_z$ from equation \eqref{eq:kz} must be superimposed. The first solution is for $+k_z$, which represents plane waves propagating (radiating) forward to $z>0$. The second solution is for $-k_z$, which represent evanescent waves which radiate into the hemisphere $z<0$. In the forward direction, these waves decay, and asymptotically approach zero magnitude well before reaching $\sim z_c$.  

In order to propagate the APWS spectrum from $z=0$ to an arbitrary measurement plane at $z \neq 0$, we multiply the spectrum $\vec{A}$ by the correct propagation factor for the direction we want to propagate. The field can then be re-created at the new plane plane with an inverse FFT:

\begin{equation} \label{eq:apws_ifft}
	\vec{E}(x,y,z) = \frac{1}{({2\pi})^2}
    \int_{-\infty}^{\infty} \int_{-\infty}^{\infty}
    \vec{A}(k_x,k_y)
    e^{\pm ik_zz}
    e^{-i(k_xx+k_yy)}
    dk_xdk_y
\end{equation}
From equation \eqref{eq:apws_ifft} we see that the APWS field (the Fourier transform of the $\vec{E}$ field in the measurement plane) is sufficient to describe the field at all points in the ($x,y,z$) coordinate space.

Figure \ref{fig:beam_prop} shows an example of amplitude-only propagation of the beam both backwards from the measurement plane at $z$=0 to the focal plane at $z$ = -33 mm, and then outwards through a series of planes. The reconstructed beam is measured in the coordinate system defined by the measurement plane. Ideally, the beam can be reconstructed at a plane of interest, such as the location of other optical elements in the system. 

\subsection{Near-Field to Far-Field Transformation}  \label{sec:introNF2FF}

 \begin{figure}
     \centering
	\includegraphics[width = 5 in]{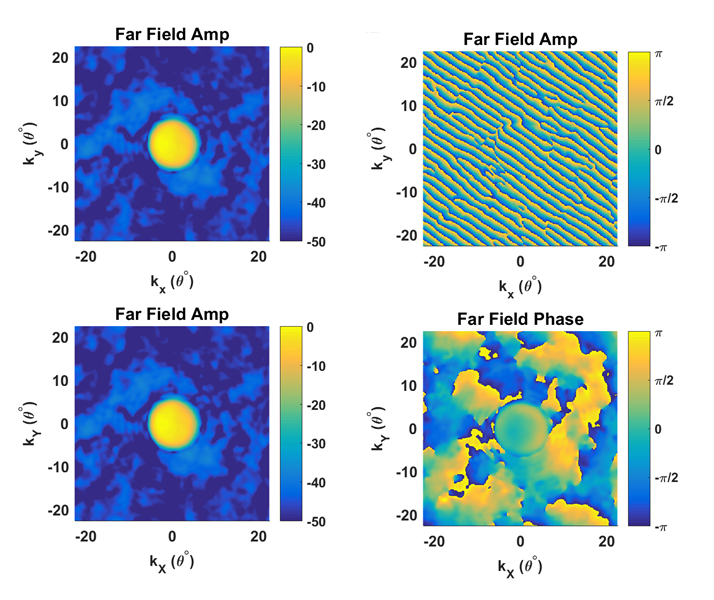}
    \caption[~~ APWS Phase Correction Example]
    {An example of applying the phase correction of equation~\ref{eq:phs_shift}
    to a far-field transformation. The right two panels show the amplitude projection
    which remains constant, and the left two panels show the phase transformation. 
    The top left panel shows the phase of the plane wave spectra as viewed edge-on,
    and the lower
    left panel shows the phase when viewed after shifting the phase angle to 
    align with the projection
    axis. }
	 \label{fig:apws_phs_shift}
\end{figure}

\begin{figure}
	\centering
	\includegraphics[width = 4.5 in]{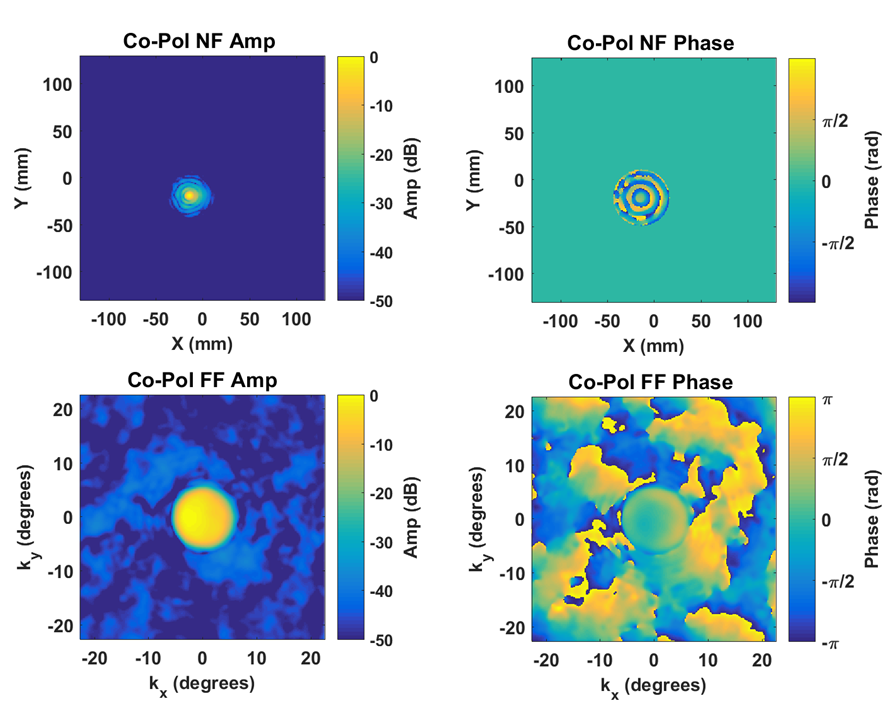}
    \caption[~~ Near-Field Spatial Masking]
    {The upper two panels of this plot show a full
    2D planar scan of a 880-pixel focal plane array. Data was recorded individually
    for each pixel, and the near-field data in amplitude (right) and phase (left)
    from a representative pixel is shown. A 2D 
    Hanning window centered on the amplitude maximum of the near-field beam is 
    applied in the upper two panels of this figure. The lower two panels show the 
    far-field transformation. Any residual diffraction is likely an artifact of the optical 
    system of the array and not from the analysis pipeline.}
	 \label{fig:apws-mask}
\end{figure}

A special case of beam propagation is to look at the beam at a plane infinitely far from the image plane. In optics terms, this is referred to as the far-field pattern, and is effectively an image of the aperture of the optical system (i.e. illumination of the secondary mirror of a telescope).  A few phase-less measurement techniques have been demonstrated to solve for the far-field antenna pattern from near-field data, but these techniques are either susceptible to finding erroneous solutions or require additional measurement planes \cite{Isernia1996-NF2FF-amponly,Tkadlec2005-NFtoFF-pointswarm}.

In the special case of being interested in propagating to the far-field at $z\to \infty $, we can take advantage of the Method of Stationary Phase to find a shorthand solution for the far-field $\vec{E}(x,y,z\to\infty)$.  A rigorous description of this method can be found in Born \& Wolfe 1994 \cite{PrincipleOptics-txtbook1994}, but is outside the scope of this analysis. The method of stationary phase makes the approximation that as the beam diverges from the near field, in the very far-field the plane waves from all other pointing directions cease to influence the point in question $(x,y,z)$. Thus at $z=\infty$, the field point is only influenced by a single plane wave from the near field. All of the other plane waves destructively interfere and cancel each other by the time they reach the far field. The effect of this approximation is that the Fourier transform of the near-field measurement  becomes the far-field radiation pattern $\vec{E}(x,y,z) \approx \vec{A}(k_x,k_y)$, which can be found using equation \eqref{eq:measplane2apws}. Figure  \ref{fig:NF2FF} shows a comparison between the near-field and APWS-fields, which are also the far-field beam patterns.   

If the near-field measurement includes data from an array instrument with many beams, the amplitude maximum of any non-central beam will not be in the center of the measurement plane.  Because of this, the transformation will propagate from the center of the scan plane rather than the center of each beam, and the initial far field transformation will show an edge-on view of a plane wave pattern which appears 'striped'. We mathematically correct for this projection by shifting the phase by the equation

\begin{equation}
	\vec{E}_{ff} = \vec{E}_{ff} * \exp{\left[i\left(k_xx_o + k_yy_o+k_zz_o\right)\right]}
\label{eq:phs_shift}
\end{equation}

where $\left(k_x,k_y,k_z\right)$ are the coordinate axis in the far field and $\left(x_o,y_o,z_o\right)$ are the coordinates of the central peak found in the near-field, with $z_o = 0$.  The effect of this phase shift is to re-center the $k$-space coordinate system at the amplitude maximum before transforming into the far field. The effect of this shift can be seen in the two right panels of figure \ref{fig:apws_phs_shift}.

Additionally, we can use spatial filtering techniques in the near-field measurements to remove truncation effects in the far-field pattern. Though a single pixel measurement will end sufficiently in the noise floor of the radiation pattern, for focal plane array measurements  significant off-axis signal, such as stray light reflections in the device substrate, can produce a diffraction pattern in the far-field transformation. We can reduce this effect by applying a circularly-symmetric spatial mask to the near-field data. Figure \ref{fig:apws-mask} shows the near-field data from a single pixel in an array with the spatial mask applied to the near-field data. The far field data has had the phase correction of \eqref{eq:phs_shift} applied. 

\section{Conclusions} \label{sec:conclusions}

The analysis presented here follows a sequential processing pipeline that can be used to characterize astronomical instruments. This is not a full description of all techniques available with complex field measurements, though it is detailed enough to characterize basic properties of the system, and are more detailed than thermal, amplitude-only beam scanning techniques. A processing pipeline using these techniques has been developed and the results are the subject of a subsequent publication.

\acknowledgments % equivalent to \section*{ACKNOWLEDGMENTS}       
 
The authors would like to thank the A-MKID collaboration for the test microwave kinetic inductance detector array that is the base data for the figures of these examples.  

% References
\bibliography{report} % bibliography data in report.bib
\bibliographystyle{spiebib} % makes bibtex use spiebib.bst

\end{document}